\documentclass{article}
\usepackage{spconf,amsmath,graphicx}
\usepackage{multirow}
\usepackage{cite}
\usepackage{booktabs}
\usepackage{bm}

\title{Improving Text-Independent Speaker Verification with Auxiliary Speakers Using Graph}
\name{Jingyu Li, Si-Ioi Ng, Tan Lee}
\address{The Chinese University of Hong Kong, Department of Electronic Engineering, Hong Kong}
\begin{document}
%
\maketitle
\begin{abstract}
The paper presents a novel approach to refining similarity scores between input utterances for robust speaker verification. Given the embeddings from a pair of input utterances, a graph model is designed to incorporate additional information from a group of embeddings representing the so-called auxiliary speakers. The relations between the input utterances and the auxiliary speakers are represented by the edges and vertices in the graph. The similarity scores are refined by iteratively updating the values of the graph’s vertices using an algorithm similar to the random walk algorithm on graphs. Through this updating process, the information of auxiliary speakers is involved in determining the relation between input utterances and hence contributing to the verification process. We propose to create a set of artificial embeddings through the model training process. Utilizing the generated embeddings as auxiliary speakers, no extra data are required for the graph model in the verification stage. The proposed model is trained in an end-to-end manner within the whole system. Experiments are carried out with the Voxceleb datasets. The results indicate that involving auxiliary speakers with graph is effective to improve speaker verification performance. 

\end{abstract}
\begin{keywords}
speaker verification, deep neural network, graph model, random walk
\end{keywords}
\section{Introduction}
\label{sec:intro}

Speaker verification (SV) systems are designed to distinguish whether an input utterance is from the claimed speaker. In earlier studies, speaker-dependent acoustic features are modeled by Gaussian Mixture Model-Universal Background Model (GMM-UBM)\cite{reynolds2000speaker} and the decision of verification is made based on estimated likelihoods. Joint Factor Analysis (JFA)\cite{kenny2007joint} and I-vector\cite{dehak2010front} consider the information in a speech utterance as the mixture of different factors, e.g., speaker and session. These factors are utilized to extract speaker-relevant information for verification. Deep neural network (DNN) embedding based models represent a predominant approach to SV in recent years\cite{variani2014deep,snyder2018x,kim2019deep,zeinali2019but,desplanques2020ecapa}. 
Speaker-relevant embeddings are extracted by a DNN model. A back-end scoring method is applied on the embeddings to measure the similarity between the test and reference utterances, e.g., cosine similarity\cite{kim2019deep,Nagrani17,xie2019utterance}, PLDA\cite{snyder2018x,jiang2019effective,liu2019large}. Metric learning methods have also been successfully applied to embedding scoring for SV \cite{chung2020defence,Nagrani19}. A high similarity between the embeddings indicates the two utterances are likely to be from the same speaker. 

Generally, back-end scoring is focused on estimating pairwise similarity between one input utterance (to be verified) and one reference utterance, without involving the information from any other utterances. However, speaker-related information provided by a single utterance is limited and may not be reliably captured in the embedding. For instance, utterances from the same speaker could be recorded under different circumstances, or a speaker may give speeches in different languages or accents. Embeddings extracted from these utterances may deviate significantly, although they come from the same speaker. Thus in one-to-one comparison with a pair of utterances, embedding based similarity score may be sub-optimal to achieve robust performance.

Incorporating additional information from other utterances into the pairwise similarity estimation is beneficial. Consider a pair of utterances $A$ and $B$ that have a low similarity score. If there exists utterance $C$ that has high similarity to both $A$ and $B$, it is probable that $A$ and $B$ are closely related or from the same speaker. Increasing the $A$-to-$B$ similarity score is expected to avoid false rejection. In this study, we define this extra utterance $C$ as from an ``auxiliary speaker”, meaning that it is supplementary and providing assistance to the goal of comparing the two target utterances. Since each speaker is represented by a single utterance in the SV process, the term ``auxiliary speaker" is made equivalent to ``auxiliary utterance" hereafter in this paper. The auxiliary speakers may carry additional information, e.g., different environment sounds, recording devices. In conventional SV setup, supplementary data are not available or not allowed. We propose the notion of ``ghost speakers'' in this paper. As the name implies, ``ghost speakers'' are not real human speakers. They refer to a set of specially prepared speaker embeddings, which are generated during model training and used as ``auxiliaries" during the verification process.


Graph structure has been widely adopted for describing the relation of data samples. With the help of DNN, deep graph network has demonstrated successful applications in many fields\cite{kipf2016semi,velivckovic2017graph,garcia2018few,bertasius2017convolutional}.
In the present study, we propose a trainable graph model, referred to as Auxiliary Speakers Graph (ASG), to incorporate the information from auxiliary utterances into the refinement of similarity score. In an ASG, similarity scores on utterance pairs are represented on the graph vertices. They are refined using the information from auxiliaries by updating the vertices' values according to the edges. The updating process is similar to the well-known random walk on graphs\cite{lovasz1993random}. Similar graph update methods have shown effectiveness in other tasks, e.g., speaker identification\cite{wang2021speaker,chen2021graph} and person re-identification\cite{shen2018person,shen2018deep}. ASG takes speaker embeddings as input and can be trained with the embedding extraction model in an end-to-end manner.

Extensive experiments are carried out on the Voxceleb1\cite{Nagrani17} and Voxceleb2\cite{Chung18b} databases to evaluate the proposed model. The ASG model shows superior performance to the baseline system that does not use auxiliary speakers. The auxiliaries can be obtained either from available speech data or model-generated embeddings, i.e., ghost speakers. Moreover, we show that ASG can outperform score normalization without training under some circumstances.

\section{Related Work}
\label{sec:related_work}

\subsection{Score normalization}
Score normalization is a commonly used approach in SV that aims to leverage speech data from other speakers \cite{shum2010unsupervised,reynolds1997comparison,auckenthaler2000score,zheng2006comparative}. The original scores of input utterances are derived from a trained model and normalized with respect to the score statistics on a cohort dataset. Our proposed approach is different from score normalization in that the graph model for score calculation and refinement can be trained jointly with the entire SV model. Score normalization is applied only to the given scores and does not affect the training of speaker models. Moreover, by using ghost speakers, cohort datasets or other additional data are not required in the proposed method.


\subsection{Graph model in SV}

In \cite{hautamaki2008text}, a graph was used to represent speaker feature distribution, and the similarity scores were computed with the graph matching algorithm. Graph matching is invariant to the rotation or uniform scaling of features, which gives stable similarity estimation for the verification. In \cite{jung2020graph}, the graph model is constructed to connect segment-wise speaker embeddings of utterance pairs. These embeddings are updated by the attention mechanism. The similarity scores are obtained by applying affine transformation on the updated embeddings in the graph. Our proposed method adopts a similar idea of connecting speaker embeddings with a graph, with the focus on score refinement with auxiliary speakers.

\section{The Proposed Method}
\label{sec:proposed_model}

The Auxiliary Speakers Graph (ASG) model takes extracted embeddings as input and refines the similarity score using the auxiliary speakers. The details of the graph structure and update will be introduced first, then the idea of ``ghost speakers'' and score calculation in evaluation will be explained.

\subsection{Graph structure}
\label{ssec:graph_structure}

\begin{figure}[t]
  \centering
  \includegraphics[width=\linewidth]{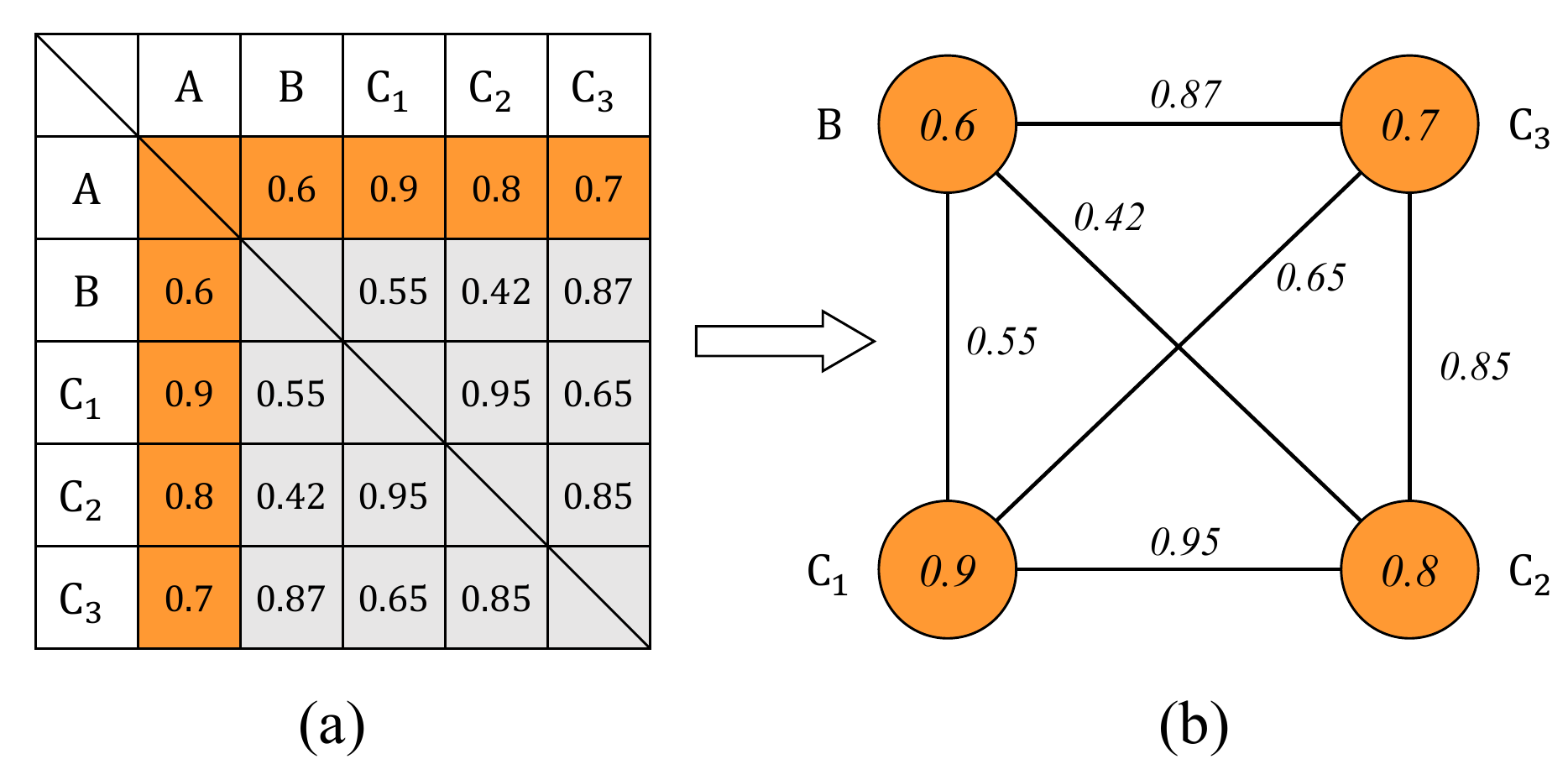}
  \vspace*{-6mm}
  \caption{(a) The matrix of similarity scores between utterances. The scores on the diagonal are ignored. (b) The ASG constructed from (a).}
  \label{fig:score_matrix}
\end{figure}

Consider a test utterance $A$, a reference utterance $B$, and $M$ auxiliaries $\{C_1, C_2,...,C_M\}$ are available for verification. Let $score_{U_1,U_2}$ denote the similarity score between utterances $U_1$ and $U_2$. Figure~\ref{fig:score_matrix} (a) shows a score matrix with $3$ auxiliaries ($M=3$). An ASG is thereby constructed on the score matrix. The ASG is an undirected graph with $N$ vertices and $N \times (N-1) / 2$ edges. The vertices are associated with the similarity scores between $A$ and other utterances, i.e., $score_{A,B} \cup score_{A,C_i}$. There are $N = M+1$ vertices in this graph. The edge connecting the vertices $score_{A,U_1}$ and $score_{A,U_2}$ is represented by $score_{U_1,U_2}$, where $U_1$ and $U_2$ could be either the reference utterance $B$ or any of the auxiliaries.

Figure~\ref{fig:score_matrix} (b) illustrates the ASG in correspondence to the score matrix in (a). The scores on vertices are given by cosine similarity between speaker embeddings. The score calculation on edges is done differently. Let $\bm{f_i}$ and $\bm{f_j}$ be the speaker embeddings extracted from $U_i$ and $U_j$ respectively and the similarity score on edge is computed by:
\begin{equation}
  S_{i,j} = Sigmoid(FC(BN((\bm{f_i}-\bm{f_j})\otimes(\bm{f_i}-\bm{f_j}))))
  \label{eq_binary_score}
\end{equation}
where $\otimes$ denotes element-wise product, $BN$ represents batch normalization layer and $FC$ stands for fully connected layer. The $Sigmoid$ function is utilized to limit the score's range in $\left( 0,1 \right)$. The scores on edges can be represented by a symmetric matrix $\bm{S} \in \mathcal{R}^{N \times N}$.

Each vertex in the ASG gives the similarity score in a pairwise comparison, thus each one of the utterances has two roles: (1) as a reference in pairwise comparison with the input, (2) as an auxiliary to other utterances, e.g., $B$ is considered as an auxiliary when refining $score_{A,C_1}$.

\subsection{Graph update}
\label{ssec:graph_update}


The value of the target score ($score_{A,B}$) is represented on a vertex, which is updated based on other vertices and the edges connecting them. When utterance $B$ shows a strong relation to utterance $C_i$, $score_{A,C_i}$ contributes more to refine the target score $score_{A,B}$. The contribution weights in the graph update are represented by a matrix $\bm{W}$ with size $N \times N$, where $N$ is the number of vertices in the ASG and $W(i,j)$ denotes the weight from vertex $j$ to $i$. $\bm{W}$ is normalized from score matrix $\bm{S}$ on edges as:
\begin{equation}
    W(i,j) =\left\{
    \begin{aligned}
    &\frac{exp(\alpha \cdot S_{i,j})}{\sum_{j=1,j\neq i}^{N}exp(\alpha \cdot S_{i,j})}&&, i\neq j& \\
    &0&&, i = j&
    \end{aligned}
    \right.
  \label{w_softmax}
\end{equation}
$\alpha$ is a trainable parameter controlling the scale of similarity scores. If the elements on the diagonal are significantly larger than others, $\bm{W}$ will approximate an identity matrix. Therefore $W(i,i)$ is set to $0$ for all $i$ in Equation~\ref{w_softmax} to avoid self-reinforcement, and there is no self-connected edge in the graph. 

The original scores on the vertices are denoted by a vector $\bm{y} \in \mathcal{R}^N$. The update of $\bm{y}$ in the $n_{th}$ iteration is implemented by matrix multiplication in a recursive way $\bm{y_{n}}=\bm{Wy_{n-1}}$, where $n$ indexes the iteration. This update process is similar to the random walk algorithm\cite{aldous1995reversible,lovasz1993random} on graphs, where $W(i,j)$ represents the transition probability from vertex $j$ to $i$, with ${\sum_i}W(i,j)=1, \forall j$. In our proposed method, the sum of contribution weights from other vertices to $i$ is set to $1$, meaning ${\sum_j}W(i,j)=1, \forall i$.

As a refinement process, the scores updated by the graph should not deviate from the original scores $\bm{y_{0}}$ too much. Thus the updating operation is formulated as:
\begin{equation}
    \bm{y_{n}}=(1-\lambda)\bm{y_{0}} + \lambda\bm{Wy_{n-1}}
    \label{score_refine}
\end{equation}
$\lambda \in [0,1]$ balances the original score and the refined score $\bm{Wy_{n-1}}$. It is determined empirically by experiments. The refined similarity score between utterances $A$ and $B$ is denoted by $\widehat{score_{A,B}}$, and located at the first element of $\bm{y_n}$.

\subsection{Ghost speakers}
\label{ssec:ghost_speakers}
During training, the auxiliary speakers are obtained by sampling from the training dataset. In the SV decision process, only a single test utterance and a reference utterance are available. If the auxiliary speakers are taken by sampling from training data, they would have low similarity scores with the test and reference utterances, due to the mismatch on speakers. Consequently, the values in $\bm{S}$ would be very small and $\bm{W}$ tends to be an average matrix. Therefore the contribution weights from different auxiliaries would be very close in updating the similarity score and the effectiveness of the proposed graph model is degraded.

Instead of collecting extra data as auxiliaries, e.g., getting a cohort set for score normalization, the idea of ghost speakers is proposed. The intuition behind introducing the so-called ghost speakers is to create auxiliaries that can be involved in the training process and used in evaluation directly. The ghost speakers are represented by a dictionary of embeddings with the size of $G \times d$, where $G$ denotes the number of ghost speakers and $d$ is the dimension of embeddings. The $G$ ghost speakers are combined with other speaker embeddings and utilized as auxiliaries in the training process. Different from other speaker embeddings which are extracted from speech data, the values of ghost speakers are initialized randomly and updated by gradient descent directly during training. Note that the ghost speakers are speaker-independent. In other words, the test utterances use the same ghost speakers as auxiliaries in the evaluation, and the ASG is constructed based on these $G$ ghost speakers.

\subsection{Calculating similarity of variable-length utterances}
\label{ssec:score_in_test}
The evaluation stage involves two variable-length input utterances, denoted as $A$ and $B$. They are divided into multiple equal-length segments and represented as $\{A_1,A_2,...,A_p\}$, $\{B_1,B_2,...,B_q\}$. To compute the score between segment $A_m$ and $B$, one graph is built among $A_m \cup \{B_1,...,B_q\}$ and $M$ auxiliaries, which gives score vector $\bm{y^m_0} \in \mathcal{R}^{q+M}$. The scores on edges are obtained as in Equation~\ref{eq_binary_score}. The edge with a very small value $S_{i,j}$ indicates that the two connected vertices have weak relation, and these redundant vertices are not considered in refining the target score. Therefore only the top $k$ values in each row of $exp(\alpha \cdot \bm{S})$ are kept in Equation~\ref{w_softmax} when calculating $\bm{W}$, and the other values are replaced by zero. The refined score after $n$ iterations of graph update is given by: 
\begin{equation}
  \widehat{score_{A,B}} =\frac{1}{p \times q}\sum_{m}^{p}\sum_{j}^{q}y^m_n(j)
  \label{score_final}
\end{equation}
$\widehat{score_{A,B}}$ is not symmetrical, i.e. $\widehat{score_{A,B}} \neq \widehat{score_{B,A}}$. The similarity between two utterances is considered to be a symmetric score, thus the similarity score for $A$ and $B$ is given by the average of $\widehat{score_{A,B}}$ and $\widehat{score_{B,A}}$ in the evaluation stage.

\section{Experiments}
\label{sec:Experiments}

\subsection{Datasets}
\label{ssec:datasets}

The datasets used in this study are Voxceleb1\cite{Nagrani17} and Voxceleb2\cite{Chung18b}. The development set of Voxceleb2 contains 1,092,009 utterances from 5,994 speakers, which are used for model training. To evaluate the performance of models, equal error rate (EER) of each model is calculated on the clean version of Voxceleb1 test set which consists of 37,611 comparison pairs formed by 4,874 utterances from 40 speakers.

\subsection{Details of implementation}
\label{ssec:implementation}

During training, an input segment with a three-second duration is randomly cropped from each training utterance. In performance evaluation, utterances are divided into four-second long, and the neighboring segments overlap with each other by 2 seconds. The audio data are transformed into 64-dimension log Mel-filterbank (FBank) coefficients using the Librosa library\cite{mcfee2015librosa}. The embedding extraction model used here is a modified version of ResNet\cite{li2020text}. The model takes Fbank as input and the output of the ``AvgPool2" layer is used as the embedding. The embedding dimension is set to $128$.

\begin{figure}[t]
  \centering
  \includegraphics[width=\linewidth]{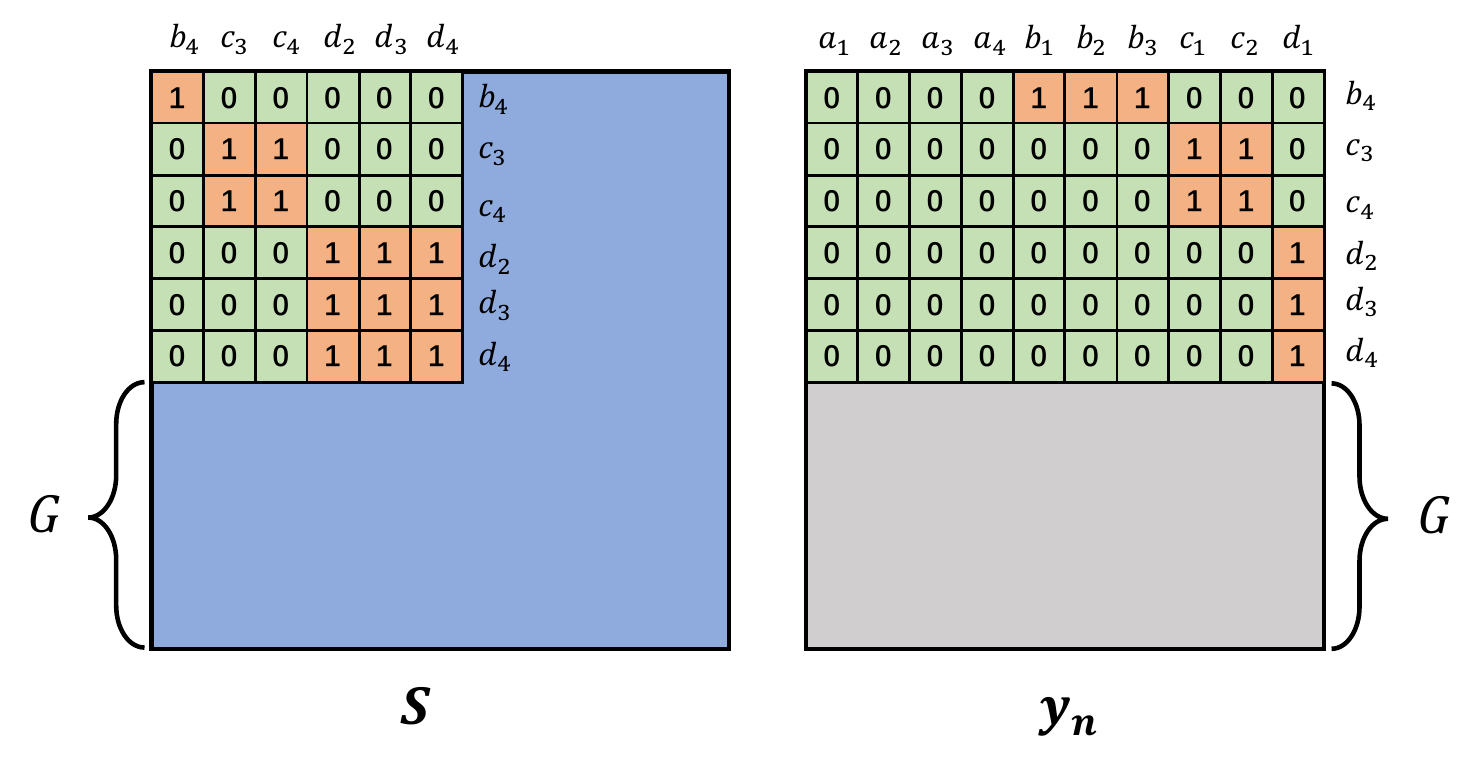}
  \vspace*{-6mm}
  \caption{Let $\{a,b,c,d\}$ represent the 4 speakers in a group, and the subscripts $\{1,2,3,4\}$ are the utterances indexes. The ``0"s and ``1"s with green or orange background indicates the supervised labels for binary cross-entropy loss. The empty part represents the scores from ghost speakers.}
  \label{fig:ladder}
\end{figure}

In each training step, four speakers are randomly sampled from the dataset and each speaker provides 4 utterances. These 16 utterances form a group and are arranged as Figure~\ref{fig:ladder} for similarity comparison. One column vector of $\bm{y_n}$ represents the vertices in one ASG after $n$ iterations of update. All graphs in $\bm{y_n}$ are updated simultaneously by multiplying with weight matrix $\bm{W}$, which is derived from the edges $\bm{S}$. The update process in one column's graph will not interfere with others. This special arrangement of sample grouping simulates the scenarios in the evaluation, where some utterances may not have any auxiliary with close relationships and some may have. For instance, utterances $\{a_1,a_2,a_3,a_4\}$ do not encounter any utterance from the same speaker in the graph, while $\{b_1,b_2,b_3\}$ will meet one similar utterance , i.e. $b_{4}$, in their graphs. There are 8 groups in one batch, giving a total of 128 utterances in a training step.

The values of $\bm{y_n}$ give the similarity scores of the corresponding utterance pairs, and a binary cross-entropy loss function is applied to it. If the two inputs come from the same speaker, the ground truth is set to $1$, else $0$. The same loss function is also applied on the edges $\bm{S}$. Combining the loss on $\bm{y_n}$ and $\bm{S}$ gives $Loss_{pair}$, but the scores on vertices and edges calculated from the ghost speakers would not be counted in this loss. The speaker embeddings are taken to predict the speaker identities, and a loss is calculated on the prediction by AM-Softmax\cite{wang2018cosface} with cross-entropy function, represented as $Loss_{id}$.

The embedding extraction model and the ASG are trained jointly for 5 epochs by the Stochastic Gradient Descent optimizer to minimize the sum of $Loss_{id}$ and $Loss_{pair}$. The weight decay is 0.001 and the momentum equals 0.9. The initial learning rate is set to 0.05 and 0.005 for the embedding extraction model and the ASG respectively. The learning rate decreases in a half-cosine schedule \cite{loshchilov2016sgdr}. No extra data augmentation is applied. After training, the utterance pairs from the test set are evaluated as in Section~\ref{ssec:score_in_test} with $k=64$ in all experiments. All model training and evaluation are implemented with PyTorch\cite{paszke2019pytorch}.

\subsection{Results}
\label{ssec:results}

The baseline performance with ResNet18 and ResNet34 give EER of $2.53\%$ and $2.21\%$ respectively. The baselines are achieved using the cosine similarity calculated between utterance embeddings without involving auxiliary speakers or graph model.


The ASG is constructed as described in Section~\ref{sec:proposed_model}, using ghost speakers as auxiliaries. The ghost speakers are trained jointly with the embedding extraction model and the graph model. The number of ghost speakers $G$ is set to $128$. The iteration of the graph update is a critical factor to the model performance. In our experiments, training with infinite iterations is found to be unstable and time-consuming. Thus we train the models at iteration $1$, $2$, and $5$. For performance evaluation, only the ghost speakers are utilized as the auxiliaries and no extra data are involved, which follows the standard SV testing procedure. Table~\ref{tab:RW_result_gen} shows the results with different settings. About $10\%$ relative improvement is observed with respect to the baselines. The iteration of graph update does not affect the performance noticeably, and $\lambda=0.2$ gives the best performance. 

In Table~\ref{tab:RW_result}, we retrain the entire system without ghost speakers. During the evaluation, $128$ or $256$ speakers are randomly sampled from the training set, their utterance embeddings are averaged on speaker basis and utilized as the auxiliaries. For each setting, the evaluation operation is repeated three times with different random seeds, and the average results are shown. Although the results are better than the baselines', they are not as good as using ghost speakers.


\begin{table}[t]
  \caption{Performances of models using  ghost speakers as auxiliaries.}
  \vspace*{4mm}
  \label{tab:RW_result_gen}
  \centering
  \begin{tabular}{lccc}
    \toprule
    \multirow{2}{*}{\shortstack[l]{\bm{$\lambda$}}}  & \multirow{2}{*}{\shortstack[l]{\textbf{Iteration}}}  & \multicolumn{2}{c}{\textbf{EER(\%)}} \\  
    \cline{3-4}
              &         &ResNet18       & ResNet34 \\
    \midrule
        \multicolumn{2}{c}{baseline} & 2.53  &  2.21  \\
    \hline
    \multirow{3}{*}{\shortstack[l]{0.1}}       & 1       & 2.34                   & 2.01  \\
                                               & 2       & 2.42                   & 2.04  \\
                                               & 5       & 2.42                   & 2.06  \\
    \hline                                           
    \multirow{3}{*}{\shortstack[l]{0.2}}       & 1       & 2.29                   & 2.00  \\
                                               & 2       & \textbf{2.11}          & 2.03  \\
                                               & 5       & 2.28                   & \textbf{1.99}  \\    
    \hline
    \multirow{3}{*}{\shortstack[l]{0.5}}       & 1       & 2.31                   & 2.08  \\
                                               & 2       & 2.35                   & 2.15  \\
                                               & 5       & 2.34                   & 2.01  \\
    \bottomrule
  \end{tabular}
\end{table}

\begin{table}[t]
  \caption{EER(\%) of models without ghost speakers. Auxiliaries are sampled from speakers in training data, and the number of auxiliaries are 128 or 256. Iteration is 2.}
  \vspace*{4mm}
  \label{tab:RW_result}
  \centering
  \begin{tabular}{ccccc}
    \toprule
    \multirow{2}{*}{\shortstack[l]{\bm{$\lambda$}}} &\multicolumn{2}{c}{ResNet18} & \multicolumn{2}{c}{ResNet34} \\  
    \cline{2-3} \cline{4-5}  
         &128             & 256           &128    & 256\\
    \hline
    0.1  & 2.43           & 2.46          & 2.11  & 2.11 \\
    0.2  & \textbf{2.35}  & \textbf{2.35} & 2.16  & 2.16 \\
    0.5  & 2.36           & 2.36          & 2.11  & \textbf{2.09} \\
    \bottomrule
  \end{tabular}
  \vspace{-2mm}
\end{table}

\begin{figure}[t]
  \centering
  \includegraphics[width=\linewidth]{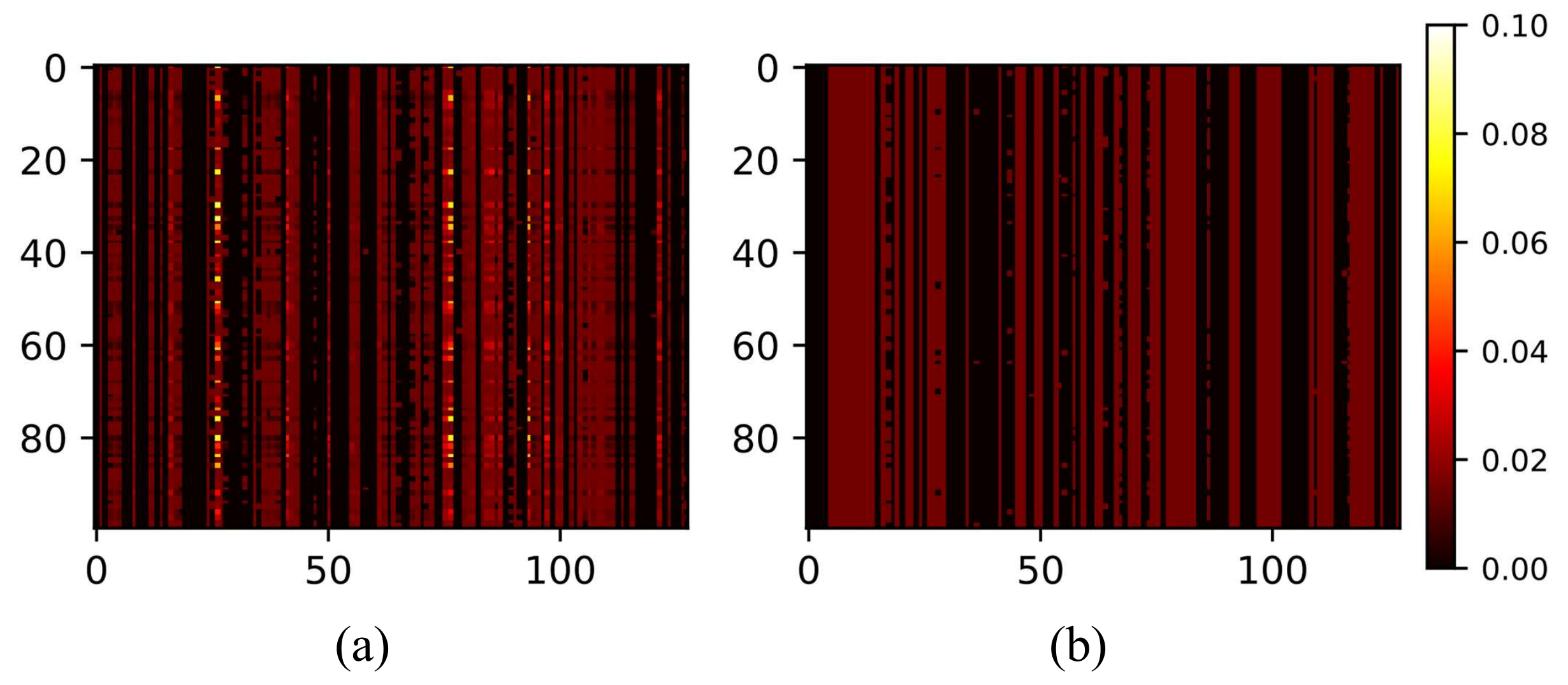}
  \vspace*{-6mm}
  \caption{(a) The contribution weights from the ghost speakers to 100 test data. (b) The contribution weights from 128 auxiliaries sampled from the training set to 100 test data.}
  \label{fig:visulization}
  \vspace{-2mm}  
\end{figure}

To better understand the different effects of using auxiliaries from the training set and using ghost speakers, 100 test utterances are randomly sampled from the test set. The auxiliaries' contribution weights $\bm{W}$ on these 100 sampled utterances are plotted as in Figure~\ref{fig:visulization}. In the weight matrix of (a), different ghost speakers show different weights to the test data. Most of the weights are very small and close in (b). The comparison between (a) and (b) suggests that different ghost speakers can contribute diverse information in the graph. This explains why using ghost speakers gives superior performance as compared with using training utterances as auxiliaries.

\subsection{Ablation study}
\label{ssec:ablation}

The proposed ASG model is similar to score normalization in that extra utterances are used to refine similarity scores. It should be noted that score normalization is a training-free method. For a fair comparison of these two methods, the calculation of graph's edges $\bm{S}$ in Equation~\ref{eq_binary_score} is replaced by the cosine similarity between embeddings, and the ghost speakers are removed. Thus the ASG model can be used for evaluation without training. First, the utterances from Voxceleb1 test set are utilized as auxiliaries. Note that using data from test set is not allowed in the standard SV process. However, in real-world applications, there may exist a large amount of unlabeled data from potential claimed speakers, and they can be utilized to improve performance. Therefore evaluation in this scenario is meaningful. Different values of $\lambda$ for graph update and $\alpha$ in Equation~\ref{w_softmax} are evaluated in the experiments. Four types of score normalization are evaluated, and the cohort set is given by the auxiliaries used in the graph.



The results are shown as in Table~\ref{tab:EER_test}. Under the assumption that all test utterances are available in the evaluation stage, the use of ASG leads to significant improvement on EER and outperforms the score normalization methods. The reason for the improvement is that the model is able to find out utterances from the potential reference speaker in the test set by selecting top $k$ similar utterances. Because the same speaker as the reference utterance must exist in the auxiliaries, the refined score is approximated as the weighted average of the scores between the test utterance and $k$ utterances from the potential reference speaker. The models are reevaluated with 1024 auxiliaries, which are randomly sampled from the test utterances. The experiments are repeated three times with different sampled auxiliaries, and the average results are reported. The performance of ASG declines noticeably, and in some cases is even worse than the baseline. Score normalization improves the baseline performance with a large margin by utilizing the utterances from the test set, and it is not much affected by the number of auxiliaries (size of cohort set). 

The performance decline of ASG is caused by missing the auxiliaries that have strong relation to the input utterances. In other words, the edges $\bm{S}$ between the reference utterance and auxiliaries are all very small, and no auxiliary is distinct. The refinement $\bm{Wy_{n-1}}$ is degraded to the average of $\bm{y_{n-1}}$. Additionally, the vertices from these weakly-connected auxiliaries should not be regarded as important for score refinement. While the edges are normalized by Equation~\ref{w_softmax}, these weakly-connected vertices in $\bm{y_{n-1}}$ still contribute $\lambda$ of the final score $\bm{y_{n}}$. 

\begin{table*}[t]
  \caption{EER(\%) of models using utterances from test set as auxiliaries. Top $k=64$, iteration is $1$.}
  \vspace*{4mm}
  \label{tab:EER_test}
  \centering
  \begin{tabular}{ll|cccccc|cccc}
    \toprule
    \multirow{2}{*}{\shortstack[l]{\textbf{Model}}} &
    \textbf{Num. of}  &  \multicolumn{2}{c}{$\alpha=1$} & \multicolumn{2}{c}{$\alpha=5$} & \multicolumn{2}{c|}{$\alpha=10$} & \multicolumn{4}{c}{\textbf{Score Norm.}}\\  
    \cline{3-12}
      & \textbf{Utt.}& $\lambda=0.5$ & $\lambda=0.8$ & $\lambda=0.5$ & $\lambda=0.8$  
      & $\lambda=0.5$ & $\lambda=0.8$   & z & t & zt & s \\
    \midrule
    \multirow{2}{*}{\shortstack[l]{\textbf{ResNet18}}}  & All   & 1.50 & \textbf{1.12} & 1.54 & 1.21 & 1.66 & 1.33 & 
                                                                2.12 & 2.10 & 2.03 & 1.92 \\ 
                                                        & 1024  & 2.44 & 3.69 & 2.43 & 2.87 & 2.14 & 2.11 & 
                                                                2.11 & 2.13 & 2.06 & \textbf{1.94} \\ 
    \hline
    \multirow{2}{*}{\shortstack[l]{\textbf{ResNet34}}}  & All   & 1.29 & \textbf{0.96} & 1.32 & 0.98 & 1.44 & 1.12 & 
                                                                1.73 & 1.83 & 1.70 & 1.56 \\ 
                                                        & 1024  & 2.39 & 3.27 & 2.07 & 2.43 & 1.81 & 1.71 & 
                                                                1.74 & 1.80 & 1.66 & \textbf{1.55} \\ 
    \bottomrule
  \end{tabular}
    \vspace*{-2mm} 
\end{table*}

\begin{table*}[t]
  \caption{EER(\%) of models with self-connected edges using speakers' embeddings from training set as auxiliaries. Top $k=512$, iteration is $1$. \textbf{SN.} stands for score normalization. In \textbf{SN.+ASG}, $\alpha=0.1$, $\lambda=0.7$.}
  \vspace*{4mm}
  \label{tab:EER_training}
  \centering
  \begin{tabular}{ll|cccccc|cccc|cc}
    \toprule
    \multirow{2}{*}{\shortstack[l]{\textbf{Model}}} &
    \textbf{Num. of}  &  \multicolumn{2}{c}{$\alpha=0.1$} & \multicolumn{2}{c}{$0.5$} & \multicolumn{2}{c|}{$1$} & \multicolumn{4}{c|}{\textbf{SN.}} & \multicolumn{2}{c}{\textbf{SN.+ASG}} \\  
    \cline{3-14}
      & \textbf{Utt.}& $\lambda=0.5$ & $0.7$ & $0.5$ & $0.7$ 
      & $0.5$ & $0.7$   & z & t & zt & s & zt & s \\
    \midrule
    \multirow{2}{*}{\shortstack[l]{\textbf{ResNet18}}}  & All  & 2.47 & 2.47 & 2.47 & 2.46 & 2.48 & 2.47 & 2.47 
                                                        & 2.51 & 2.49 & 2.49 & \textbf{2.37} & 2.42 \\ 
                                                        & 1024 & 2.50 & 2.47 & 2.49 & 2.47 & 2.49  & 2.47 
                                                        & 2.53 & 2.55 & 2.51 & 2.51 
                                                        & \textbf{2.44} & \textbf{2.44} \\ 
    \hline
    \multirow{2}{*}{\shortstack[l]{\textbf{ResNet34}}}  & All  & 2.15 & 2.15 & 2.16 & 2.14 & 2.15 & 2.15
                                                        & 2.16 & 2.16 & 2.13 & 2.18 & \textbf{2.09} & 2.14 \\ 
                                                        & 1024  & 2.19 & 2.17 & 2.19 & 2.17 & 2.19 & 2.18 & 2.17 & 2.18 & 2.14 & 2.16 & \textbf{2.09} & 2.13 \\ 
    \bottomrule
  \end{tabular}
    \vspace*{-2mm} 
\end{table*}

Self-connected edges are added in ASG as $S_{i,i}=1$ in the next experiment. The self-connected edges tend to dominate the graph update weights $\bm{W}$ and suppress the function of other vertices in $\bm{y_{n-1}}$ when the edges $\bm{S}$ between the reference utterance and auxiliaries are all very small. The models are evaluated utilizing the speaker-wise average of the embeddings in Voxceleb2 training set as auxiliaries, which gives 5994 auxiliaries in total. Utilizing a larger $k$ is founded useful. We report the results in Table~\ref{tab:EER_training} under two conditions, using all auxiliaries or 1024. Using ASG gives little performance improvement compared with the baseline, and is close to the score normalization. ASG can be applied after score normalization, which further improves the performance. The trained ASG in Section~\ref{ssec:results} uses fewer auxiliaries and achieves lower EER than the ASG with untrained edges. This indicates that the trained edges can learn better information to update the graph.

\begin{table}[t]
  \caption{EER(\%) of ECAPA-TDNN}
  \vspace*{4mm}
  \label{tab:ecapa}
  \centering
  \begin{tabular}{l|cc}
    \toprule
                       & ECAPA-512       & ECAPA-1024          \\  
    \hline              
    baseline           & 1.27            & 1.05                \\
    \hline
    ASG                & 1.27            & 1.05                \\
    s-norm             & 1.20            & 0.99                \\
    zt-norm            &   1.17          & 1.01                \\
    s-norm + ASG       & \textbf{1.12}   & \textbf{0.97}       \\
    zt-norm + ASG      & 1.16            & \textbf{0.97}       \\
    \bottomrule
  \end{tabular}
  \vspace*{-2mm}  
\end{table}

\subsection{On other models}
\label{ssec:sota}
We apply the untrained ASG in Section~\ref{ssec:ablation} on ECAPA-TDNN\cite{desplanques2020ecapa} with 512 and 1024 channels. ECAPA-TDNN is trained on Voxceleb2 training set. Each utterance is augmented three times using MUSAN dataset\cite{snyder2015musan} and RIR dataset\cite{ko2017study}. Other augmentation methods in \cite{desplanques2020ecapa}, like tempo up/down or SpecAugment\cite{park2019specaugment}, are not used. Each ECAPA-TDNN is trained for 20 epochs. The learning rate is initialized as 0.001 and decayed by a factor of 0.1 every 8 epochs.

The results are shown as in Table~\ref{tab:ecapa}. The performance keeps unchanged when only ASG is used, and the zt-norm and s-norm\cite{shum2010unsupervised} give slight improvement. Because ASG and score normalization refine the scores in different manners, we can also employ the graph update on the normalized score. Applying these two methods sequentially gives the best performance.

\section{Conclusions and Future Works}
\label{sec:conclusion}

In this paper, we investigate the effect of utilizing auxiliary speakers with a graph model in speaker verification. The proposed Auxiliary Speakers Graph (ASG) model represents the relation between utterances by edges and refines the similarity scores on the vertices in a manner similar to random walk. ASG can be trained with the embedding extraction model in an end-to-end manner. Incorporating ``ghost speakers" as auxiliaries, ASG improves the performance without requiring additional data in the evaluation stage. The ablation study shows that ASG can also be utilized as a training-free refinement method and outperform score normalization under certain circumstances.

The choice of auxiliary speakers and the parameters in ASG affect the performance greatly. Further investigations on these factors are needed. The information of speech production is not considered in the generation of ghost speakers, and thus the ghost speakers may not match the patterns of embeddings extracted from real speech data. In future work, generative adversarial network (GAN)\cite{goodfellow2014generative} or variational auto-encoder (VAE)\cite{kingma2013auto} could be applied to generating the ghost speakers and producing more robust embeddings.

\section{Acknowledgements}
\label{sec:acknowledgements}
The first author was supported by the Hong Kong Ph.D. Fellowship Scheme.

\bibliographystyle{IEEEbib}
\bibliography{strings,refs}

\end{document}